\documentclass[twocolumn,english,aps,prl,superscriptaddress,raggedbottom,showpacs,floatfix,longbibliography]{revtex4-1}
\usepackage[T1]{fontenc}
\usepackage[utf8]{inputenc}
\usepackage{color}
\usepackage{babel}
\usepackage{verbatim}
\usepackage{amsmath}
\usepackage{amssymb}
\usepackage{graphicx}
\usepackage{esint}
\usepackage{soul}
\usepackage{hyperref}
\usepackage{natbib}

\makeatletter
\@ifundefined{textcolor}{}
{%
 \definecolor{BLACK}{gray}{0}
 \definecolor{WHITE}{gray}{1}
 \definecolor{RED}{rgb}{1,0,0}
 \definecolor{GREEN}{rgb}{0,1,0}
 \definecolor{BLUE}{rgb}{0,0,1}
 \definecolor{CYAN}{cmyk}{1,0,0,0}
 \definecolor{MAGENTA}{cmyk}{0,1,0,0}
 \definecolor{YELLOW}{cmyk}{0,0,1,0}
}

\@ifundefined{definecolor}
 {\usepackage{color}}{}
\usepackage{babel}

\makeatother

\begin{document}

\title{Magnetic end-states in a strongly-interacting one-dimensional topological Kondo insulator}

\author{Alejandro M. Lobos}
\email{lobos@ifir-conicet.gov.ar}

\affiliation{Facultad de Ciencias Exactas Ingenier\'ia y Agrimensura, Universidad
Nacional de Rosario and Instituto de F\'isica Rosario,  Bv. 27 de Febrero
210 bis, 2000 Rosario, Argentina}

\affiliation{Condensed Matter Theory Center and Joint Quantum Institute, Department
of Physics, University of Maryland, College Park, Maryland 20742-4111,
USA}

\author{Ariel O. Dobry}

\affiliation{Facultad de Ciencias Exactas Ingenier\'ia y Agrimensura, Universidad
Nacional de Rosario and Instituto de F\'isica Rosario,  Bv. 27 de Febrero
210 bis, 2000 Rosario, Argentina}

\author{Victor Galitski}

\affiliation{Condensed Matter Theory Center and Joint Quantum Institute, Department
of Physics, University of Maryland, College Park, Maryland 20742-4111,
USA}

\affiliation{School of Physics, Monash University, Melbourne, Victoria 3800, Australia}

\date{\today}
\begin{abstract}
Topological Kondo insulators are strongly correlated materials, where
itinerant electrons hybridize with localized spins giving rise to
a topologically non-trivial band structure. Here we use non-perturbative
bosonization and renormalization group techniques to study theoretically
a one-dimensional topological Kondo insulator, described as
a Kondo-Heisenberg model where the Heisenberg spin-1/2 chain is coupled
to a Hubbard chain through a Kondo exchange interaction in the $p$-wave
channel (i.e., a strongly correlated version of the prototypical Tamm-Schockley
model). We derive and solve renormalization group equations at two-loop
order in the Kondo parameter, and find that, at half-filling, the
charge degrees of freedom in the Hubbard chain acquire a Mott gap,
even in the case of a non-interacting conduction band (Hubbard parameter
$U=0$). Furthermore, at low enough temperatures, the system maps
onto a spin-1/2 ladder with local ferromagnetic interactions along
the rungs, effectively locking the spin degrees of freedom into a
spin-$1$ chain with frozen charge degrees of freedom. This structure
behaves as a spin-1 Haldane chain, a prototypical interacting topological
spin model, and features two magnetic spin-$1/2$ end states for chains
with open boundary conditions. Our analysis allows to derive an insightful
connection between topological Kondo insulators in one spatial dimension
and the well-known physics of the Haldane chain, showing that the
ground state of the former is qualitatively different from the predictions
of the na{\"i}ve mean-field theory.
\end{abstract}
\maketitle

\section{Introduction}

Starting with the pioneering works of Kane and Mele \cite{Kane05_Z2_Topological_invariant_in_QSHE,Kane05_Kane_Mele_model_graphene}
and others \cite{Moore07_Topological_invariants_in_time_reversal_bands,Roy09_Z2_invariants_for_QSH_systems,Fu07_Topological_insulators_in_3D},
there has been a surge of interest in topological characterization
of insulating states \cite{Hasan10_Topological_insulators_review,Qi11_Review_TI_and_TSC,Bernevig_book_TI_TSC}.
It is now understood that there exist distinct symmetry-protected
classes of non-interacting insulators, such that two representatives
from different classes can not be adiabatically transformed into one
another (without closing the insulating gap and breaking the underlying
symmetry along the way). A complete topological classification of
such band insulators has been developed in the form of a ``periodic
table of topological insulators'' \cite{Kitaev_TI_classification,Ryu10_Topological_classification}.
Furthermore, it was realized that the non-trivial (topological) insulators
from this table possess, as their hallmark features, gapless boundary
modes. The latter have been spectacularly observed in a variety of
experiments in both three \cite{Hsieh08_Topological_Dirac_insulator_in_QSH_phase,Xia09_Topological_insulator_in_Bi2Se3}
and two-dimensional systems \cite{Konig07_Observation_of_QSH_in_HgTe,Knez11_Evidence_of_edge_modes_in_InAsGaSb_QSHI,Nowack13_Edge_modes_in_QSHI_by_scanning_SQUID,Spanton14_Edge_modes_in_QSHI_by_scanning_SQUID}.

The aforementioned classification however is limited to non-interacting
systems and as such it represents a classification of single-particle
band structures. Adding interactions to the theory leads to significant
complications. To understand and classify strongly-interacting topological
insulator phases in many-particle systems is a fundamental open problem
in condensed matter.

\begin{figure}
\includegraphics[bb=120bp 190bp 560bp 360bp,clip,scale=0.55]{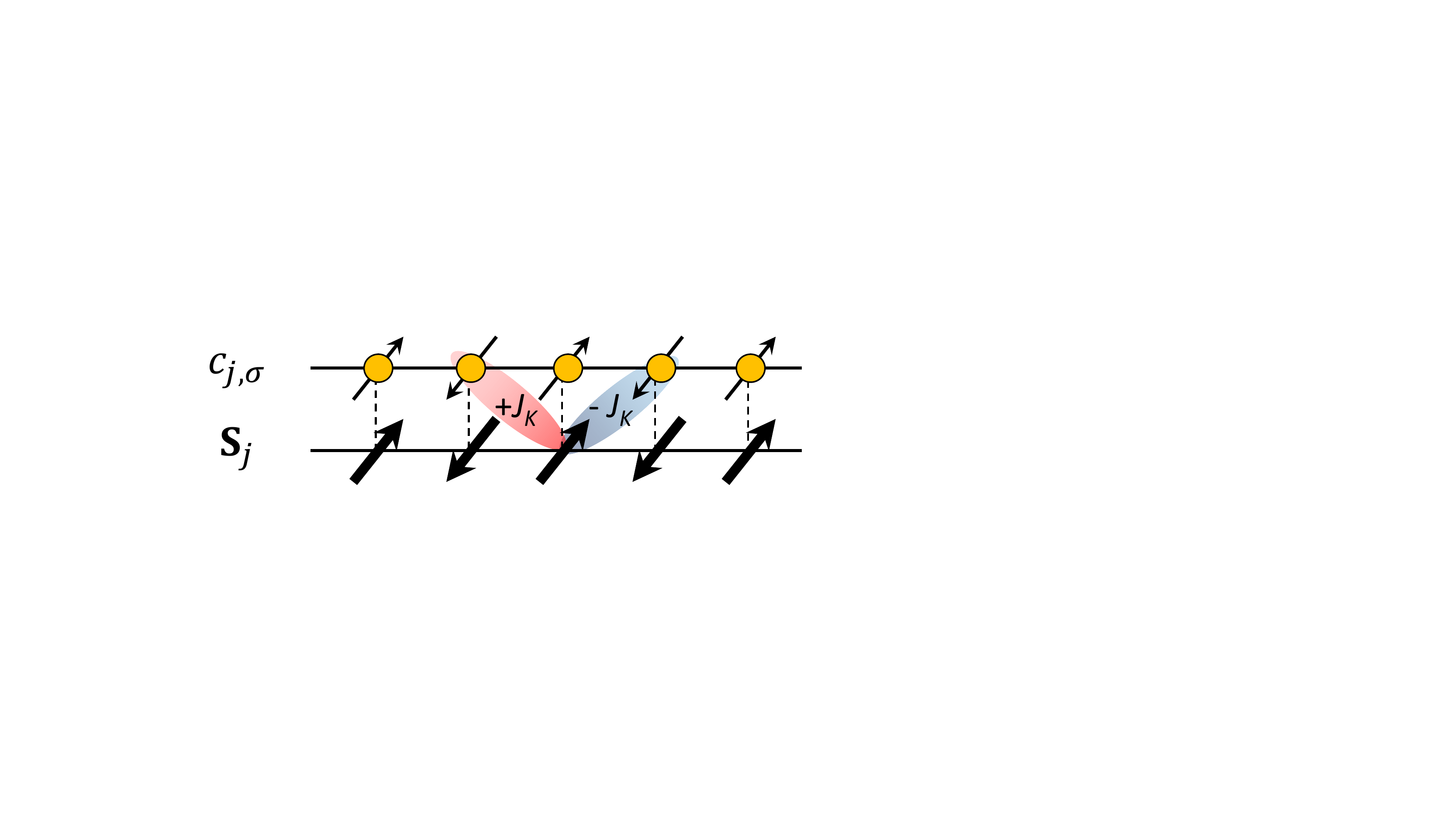}
\protect\caption{\label{system}Schematic representation of the 1D ``p-wave'' Kondo-Heisenberg
model. The top chain corresponds to the Hubbard model and the bottom
chain is the $S=1/2$ antiferromagnetic Heisenberg model. The Kondo
exchange (depicted pictorially as slanted bonds with odd inversion
symmetry) is a non-local nearest-neighbor antiferromagnetic interaction
$J_{K}>0$ which couples a spin $\mathbf{S}_{j}$ in the Heisenberg
chain with the ``p-wave'' spin-density $\boldsymbol{\pi}_{j}$ in
the Hubbard chain {[}see Eq. (\ref{eq:pi_definition}){]}. }
\end{figure}

A class of material that combines strong interactions and non-trivial
topology of emergent bands are topological Kondo insulators (TKIs) \cite{Dzero10_Topological_Kondo_Insulators}. A basic
model of these heavy fermion systems involves even-parity conduction
electrons hybridizing with strongly correlated $f$-electrons. At low
temperatures, a hybridization gap opens up and an insulating state
can be formed. Its simplified mean-field description makes it amenable
to a topological classification according to the non-interacting theory,
and a topologically-non-trivial state appears due to the opposite
parities of the states being hybridized. Although the mean-field
description (formally well-controlled in the large-$N$ approximation
 \cite{read83,coleman87,newns87}) does appear to correctly describe
the nature of the topological Kondo insulating states observed in
bulk materials so far \cite{TKI_Reich,TKI_exp1,TKI_exp2,TKI_exp3,TKI_exp4},
it is interesting to see if non-perturbative effects beyond mean-field
can qualitatively change the mean-field picture.

In contrast to higher dimensions, where reliable theoretical techniques
to treat strong interactions are scarce, there exists a rich arsenal
of such non-perturbative methods for one-dimensional systems, where
strongly-correlated ``non-mean-field'' ground states abound. Since
the Kondo insulating Hamiltonian and its mean-field treatment are
largely dimension-independent, it is interesting to consider the one-dimensional
such model as a natural playground to study interplay between strong
interactions and non-trivial topology.

With this motivation in mind, we study here a strongly-interacting
model of a one-dimensional topological Kondo insulator, i.e., a ``$p$-wave''
Kondo-Heisenberg model, introduced earlier by Alexandrov and Coleman
 \cite{Alexandrov2014_End_states_in_1DTKI}, who treated the problem
in the mean-field approximation. Here, we go beyond the mean-field
level and consider quantum fluctuations non-peturbatively using the
Abelian bosonization technique. It is shown that a ``topological
coupling'' between the electrons in the Hubbard chain and spins in
the Heisenberg chain, gives rise to a charge gap at half-filling in
the former. The relevant interaction between the remaining spin-$1/2$
degrees of freedom in the chains is effectively ferromagnetic, which
locks them into a state qualitatively similar to the Haldane's spin-$1$
chain. The ground state therefore is a strongly-correlated topological
insulator, which exhibits neutral spin-$1/2$ end modes.

While our main motivation is essentially theoretical (i.e., to allow a deeper understanding of strongly interacting topological matter), we believe our results might have direct application in ultracold atom experiments, where double-well optical superlattices loaded with atoms in $s$ and $p$ orbitals have been realized \cite{Wirth11_Superfluidity_in_p_band_optical_lattice, Soltan-Panahi11_QPT_in_multiorbital_optical_lattices}. In addition, our work might have some relevance in recent experimental results \cite{Weird_TKI_exp1,Weird_TKI_exp2,Weird_TKI_exp3}, which suggest the existence of a ferromagnetic phase transition and/or suppressed surface charge transport in samples of Samarium hexaboride ($\text{SmB}_{6}$ -- a three-dimensional topological Kondo insulator).

This article is organized as follows: in Section \ref{sec:model}
we specify the model for a 1D TKI and introduce the Abelian bosonization
description. In Section \ref{sec:RG} we present the renormalization
group analysis and discuss the quantum phase diagram of the system.
In Section \ref{sec:topo} we analyze the topological aspects of the
problem and explain the emergence of topologically protected magnetic
edge-states and in Section \ref{sec:conclusions} we present a summary
and discussion of results. Finally, in the Appendix \ref{sec:appendix_RG}
we present the technical derivation of the renormalization group equations.

\section{\label{sec:model}Model}

We start our theoretical description by considering the Hamiltonian
of the system depicted in Fig. \ref{system}, $H=H_{1}+H_{2}+H_{K}$
, where

\begin{alignat}{1}
H_{1} & =-t\sum_{j=1,\sigma}^{N_{s}-1}\left(c_{j,\sigma}^{\dagger}c_{j+1,\sigma}+\mbox{H.c.}\right)-\mu\sum_{j=1,\sigma}^{N_{s}}n_{j,\sigma}\nonumber \\
 & +U\sum_{j=1}^{N_{s}}\left(n_{j\uparrow}-\frac{1}{2}\right)\left(n_{j\downarrow}-\frac{1}{2}\right)\label{eq:1D_Hubbard}
\end{alignat}
is a fermionic 1D Hubbard chain with $N_{s}$ sites, where $n_{j,\sigma}=c_{j,\sigma}^{\dagger}c_{j,\sigma}$
is the density of spin-$\sigma$ electrons at site $j$, $\mu$ is
the chemical potential, and $U$ is the Hubbard interaction parameter.
In this work we will only focus on the half-filled case $\mu=0$,
where there is one electron per site. However, we expect our results
to remain also valid for small deviations of half-filling. The spin
chain is described by the spin-1/2 Heisenberg model

\begin{alignat}{1}
H_{2} & =J\sum_{j=1}^{N_{s}-1}\mathbf{S}_{j}\cdot\mathbf{S}_{j+1},\label{eq:1D_Heisenberg}
\end{alignat}
with $J>0$. Here we assume the same lattice parameter $a$ for both
chains $H_{1}$ and $H_{2}$. Finally, motivated by the work by Alexandrov
and Coleman  \cite{Alexandrov2014_End_states_in_1DTKI}, we assume
the following exchange coupling between the two chains

\begin{alignat}{1}
H_{K} & =J_{K}\sum_{j=1}^{N_{s}}\mathbf{S}_{j}\cdot\boldsymbol{\pi}_{j},\label{eq:H_K}
\end{alignat}
where $J_{K}>0$ is the Kondo interaction between the $j-$th spin
($\mathbf{S}_{j}$) in the Heisenberg chain and the \textit{p-wave
spin density} in the fermionic chain at site $j$, defined as

\begin{alignat}{1}
\boldsymbol{\pi}_{j} & \equiv p_{j,\alpha}^{\dagger}\left(\frac{\boldsymbol{\sigma}_{\alpha\beta}}{2}\right)p_{j,\beta},\label{eq:pi_definition}
\end{alignat}
where $p_{j,\alpha}\equiv\left(c_{j+1,\alpha}-c_{j-1,\alpha}\right)/\sqrt{2}$
is a linear combination of orbitals with $p$-wave symmetry, and $\boldsymbol{\sigma}_{\alpha\beta}$
is the vector of Pauli matrices. This model can be regarded as a strongly
interacting version of the Tamm-Shockley model  \cite{Tamm32_Tamm_model,Shockley39_Shockley_model,Pershoguba12_Surface_states}. While for our present purposes, this is an interesting ``toy model'' Hamiltonian that allows to extract a useful insight into strongly interacting topological phases, it could in principle be realized in ultracold-atom experiments [see Sec.\ref{sec:conclusions}) for details].
In the absence of interactions in the fermionic chain (i.e., $U=0$)
and in the large$-N$ mean-field approximation, Alexandrov and Coleman
have shown the emergence of topologically-protected edge states arising
from the non-trivial form of the Kondo term (\ref{eq:H_K}) \cite{Alexandrov2014_End_states_in_1DTKI}.
In their mean-field approach, the effective description of the system
corresponds to non-interacting quasiparticles filling a strongly renormalized
valence band with a non-trivial topology, stemming from the charge-conjugation,
time-reversal and charge U(1) symmetry of the effectively non-interacting
Hamiltonian (see also Ref. \cite{Li13_Topological_interacting_ladders}
for a discussion of a closely related system).

In this paper, our goal is to understand the emergence of topologically
protected edge-states without introducing any decoupling of the Kondo
interaction, including the interacting case, $U\neq0$. We consider
the case of small $U$ and $J_{K}$. This is formally represented
by linearizing the non-interacting spectrum $\epsilon_{k}=-2t\cos ka$
in the fermionic chain $H_{1}$ around the Fermi energy $\mu=0$,
and taking the continuum limit where the lattice constant $a\rightarrow0$.
Then, the fermionic operators admit the low-energy representation
 \cite{giamarchi_book_1d,gogolin_1dbook}

\begin{align}
\frac{c_{j,\sigma}}{\sqrt{a}} & \sim e^{ik_{F}x_{j}}R_{1,\sigma}\left(x_{j}\right)+e^{-ik_{F}x_{j}}L_{1,\sigma}\left(x_{j}\right),\label{eq:linearization}
\end{align}
where $R_{1,\sigma}\left(x\right)$ and $L_{1,\sigma}\left(x\right)$
are right and left moving fermionic field operators, which vary slowly
on the scale of $a$. As we are interested in the edge-state physics,
we consider open boundary conditions $c_{0,\sigma}=c_{N_{s}+1,\sigma}=0$,
leading to the following constraints:

\begin{alignat}{1}
L_{1,\sigma}\left(0\right) & =-R_{1,\sigma}\left(0\right),\label{eq:constraint_x0}\\
L_{1,\sigma}\left(L_c\right) & =-e^{i2k_{F}L_c}R_{1,\sigma}\left(L_c\right),\label{eq:constrain_xL}
\end{alignat}
where $L_c=N_{s}a$ is the length of the chain. We next introduce the Abelian bosonization formalism
 \cite{giamarchi_book_1d,gogolin_1dbook}

\begin{align}
R_{1,\sigma}\left(x\right) & =\frac{F_{1,\sigma}}{\sqrt{2\pi\alpha}}e^{-i\phi_{1,R,\sigma}\left(x\right)},\nonumber \\
L_{1,\sigma}\left(x\right) & =\frac{F_{1,\sigma}}{\sqrt{2\pi\alpha}}e^{i\phi_{1,L,\sigma}\left(x\right)},\label{eq:bosonizationRL}
\end{align}
where $\alpha$ is a short distance cutoff in the bosonization procedure
(we will take $\alpha=a$ hereafter). In Eq. (\ref{eq:bosonizationRL})
$\phi_{1,\lambda\sigma}\left(x\right)$ (with $\lambda=\left\{ R,L\right\} $)
are bosonic fields obeying the commutation relations $\left[\phi_{1,R\sigma}\left(x\right),\phi_{1,R\sigma^{\prime}}\left(y\right)\right]=i\pi\text{sign}\left(x-y\right)\delta_{\sigma,\sigma^{\prime}}$,
$\left[\phi_{1,L\sigma}\left(x\right),\phi_{1,L\sigma^{\prime}}\left(y\right)\right]=-i\pi\text{sign}\left(x-y\right)\delta_{\sigma,\sigma^{\prime}}$,
and $F_{1,\sigma}$ are Klein operators which obey anticommutation
relations $\left\{ F_{1,\sigma},F_{1,\sigma^{\prime}}\right\} =\delta_{\sigma,\sigma^{\prime}}$,
and therefore ensure the correct anticommutation relations for
fermions. Due to the constraints (\ref{eq:constraint_x0})
and (\ref{eq:constrain_xL}) introduced by the open boundary conditions,
the right and left movers are not independent, and obey the constraints

\begin{align}
\phi_{1,L,\sigma}\left(0\right) & =-\phi_{1,R,\sigma}\left(0\right)+\pi,\label{eq:boundary_condition_0}\\
\phi_{1,L,\sigma}\left(L_c\right) & =-\phi_{1,R,\sigma}\left(L_c\right)+2k_{F}L_c-\pi+2q_{\sigma}\pi.\label{eq:boundary_condition_L}
\end{align}
Here, $q_{\sigma}$ is an integer representing the occupation of the
``zero-mode'' excitations, i.e., particle-hole excitations with momentum $k=0$ and total spin $\sigma$. Its presence in Eq. (\ref{eq:boundary_condition_L}) can be understood recalling that the expression of the non-chiral bosonic field $\phi_{1,\sigma}=\left(\phi_{1,R,\sigma} +\phi_{1,L,\sigma}\right)/2$ is \cite{lecheminant02_magnet} 
\begin{align*} \phi_{1,\sigma}\left(x\right)&=\frac{\pi}{2}+\left( k_F L_c - \pi +\pi q_\sigma \right)\frac{x}{L_c} \nonumber\\ &+ \sum_{n=1}^\infty \frac{\sin \left(k_n x \right)}{\sqrt{n}} \left( \alpha_{n,\sigma} + \alpha^\dagger_{n,\sigma} \right),
\end{align*} where $k_n\equiv \frac{\pi n}{L_c}$ with integer $n>0$,  and $\alpha^\dagger_{n,\sigma}$ are bosonic operators obeying the commutation relation $\left[\alpha_{n,\sigma},\alpha^\dagger_{m,\sigma^\prime}\right]=\delta_{n,m}\delta_{\sigma,\sigma^\prime}$ (see Refs. \onlinecite{delft_bosonization,lecheminant02_magnet} for details). From here we obtain the additional commutation relations
 \cite{lecheminant02_magnet}

\begin{align}
\left[\phi_{1,R\sigma}\left(x\right),\phi_{1,L\sigma^{\prime}}\left(y\right)\right] & =\begin{cases}
-i\pi\delta_{\sigma,\sigma^{\prime}} & \text{for }0<x,y<L_c,\\
0 & \text{for }x=y=0,\\
-2\pi i\delta_{\sigma,\sigma^{\prime}} & \text{for }x=y=L_c.
\end{cases}\label{eq:RL_commutation_relation}
\end{align}

The bosonization procedure applied to the 1D Hubbard model is standard
and we refer the reader to textbooks for details  \cite{giamarchi_book_1d,gogolin_1dbook}.
Introducing charge and spin bosonic fields $\phi_{1,\lambda\sigma}=\frac{1}{\sqrt{2}}\left[\phi_{1c}-\lambda\theta_{1c}+\sigma\left(\phi_{1s}-\lambda\theta_{1s}\right)\right]$
(where the convention of signs $\lambda=\left\{ R,L\right\} =\left\{ +,-\right\} $ and $\sigma=\left\{ \uparrow,\downarrow\right\} =\left\{ +,-\right\} $ is implied),
the 1D Hubbard model at half-filling (i.e., $k_{F}=\pi/2a$) becomes
 \cite{giamarchi_book_1d}

\begin{align}
H_{1} & =\sum_{\nu=c,s}\int_{0}^{L_c}dx\;\left[\frac{v_{1\nu}}{2\pi K_{1\nu}}\left(\partial_{x}\phi_{1\nu}\right)^{2}+\frac{v_{1\nu}K_{1\nu}}{2\pi}\left(\partial_{x}\theta_{1\nu}\right)^{2}\right]\nonumber \\
 & -\frac{U}{2\left(\pi a\right)^{2}}\int_{0}^{L_c}dx\;\left[\cos\left(\sqrt{8}\phi_{1c}\right)-\cos\left(\sqrt{8}\phi_{1s}\right)\right],\label{eq:1D_Hubbard_bosonic}
\end{align}
where the new fields obey the boundary conditions
\begin{align}
\phi_{1s}\left(0\right)=0 & ,\quad\phi_{1s}\left(L_c\right)=\frac{\pi}{\sqrt{2}}\left(q_{\uparrow}-q_{\downarrow}\right),\label{eq:constraint_phi1s}\\
\phi_{1c}\left(0\right)=\frac{\pi}{\sqrt{2}} & ,\quad\phi_{1c}\left(L_c\right)=\sqrt{2}\left(k_{F}L_c-\frac{\pi}{2}\right)+\frac{\pi}{\sqrt{2}}\left(q_{\uparrow}+q_{\downarrow}\right), \label{eq:constraint_phi1c}
\end{align}
and the commutation relation $\left[\phi_{1,\nu}\left(x\right),\theta_{1,\nu^\prime}\left(y\right)\right]=-i\frac{\pi}{2}\delta_{\nu,\nu^\prime} \text{sign}\left(x-y\right)$. From here, we conclude that the field $\frac{1}{\pi}\partial_x \theta_{1,\nu}\left(x\right)$ is the momentum canonically conjugated to $\phi_{1,\nu}\left(x\right)$.

As is well-known, in 1D charge and spin excitations generally decouple
and the above Hamiltonian can be split as $H_{1}=H_{1c}+H_{1s}$,
with the first line describing independent Luttinger liquids for the
charge and spin sectors, which are characterized by charge (spin)
acoustic modes with velocities $v_{1c}=v_{F}\sqrt{1+Ua/\left(\pi v_{F}\right)}$
$\left(v_{1s}=v_{F}\right)$, and Luttinger parameter controlling
the decay of correlation functions $K_{1c}=1/\sqrt{1+Ua/\left(\pi v_{F}\right)}$
($K_{1s}=1$) . The presence of the cosine terms in the second line
of (\ref{eq:1D_Hubbard_bosonic}) changes the physics qualitatively.
In the present work, we restrict our focus to the case $U\geq0$,
where the term $\sim\cos\left(\sqrt{8}\phi_{1c}\right)$ is marginally
relevant in the renormalization-group sense, and opens a Mott gap
in the charge sector. At the same time, the term $\sim\cos\left(\sqrt{8}\phi_{1s}\right)$
is marginally irrelevant at the SU(2) symmetric point, and the spin
sector remains gapless  \cite{gogolin_1dbook,giamarchi_book_1d}.%

The bosonization of the Heisenberg chain $H_{2}$ is also quite standard
and we refer the reader to the {above-mentioned} textbooks \cite{giamarchi_book_1d,gogolin_1dbook}.
A usual trick consists in representing the spin operators $\mathbf{S}_{j}$
by auxiliary fermionic operators in a half-filled Hubbard model with
interaction parameter $U^{\prime}\gg U$. Therefore, while technically
the procedure is identical to Eq. (\ref{eq:1D_Hubbard_bosonic}),
the charge degrees of freedom in the bosonized Hamiltonian, $H_{2}$,
can be assumed to be absent at the relevant energy scales of the problem
due to the Mott gap $\sim U^{\prime}$. Then, ignoring the charge
degrees of freedom and irrelevant operators in Eq. (\ref{eq:1D_Hubbard_bosonic}),
and replacing the chain label $1\rightarrow2$, we obtain

\begin{alignat}{1}
H_{2} & =\frac{v_{2s}}{2\pi}\int_{0}^{L_c}dx\;\left[\left(\partial_{x}\phi_{2s}\right)^{2}+\left(\partial_{x}\theta_{2s}\right)^{2}\right].\label{eq:H2_LL}
\end{alignat}

Finally, we bosonize the Kondo Hamiltonian. The $p$-wave spin density
in the fermionic chain and the spin density in the Heisenberg chain
are, respectively

\begin{alignat}{1}
\frac{\boldsymbol{\pi}_{j}}{a} & \sim2\left[\mathbf{J}_{1R}\left(x_{j}\right)+\mathbf{J}_{1L}\left(x_{j}\right)-\left(-1\right)^{j}\mathbf{N}_{1}\left(x_{j}\right)\right],\label{eq:pi_density}\\
\frac{\mathbf{S}_{j}}{a} & \sim\mathbf{J}_{2R}\left(x_{j}\right)+\mathbf{J}_{2L}\left(x_{j}\right)+\left(-1\right)^{j}\mathbf{N}_{2}\left(x_{j}\right),\label{eq:S_density}
\end{alignat}
where $\mathbf{J}_{aR}\left(x\right)=R_{a,\alpha}^{\dagger}\left(x\right)\left(\frac{\boldsymbol{\sigma}_{\alpha,\beta}}{2}\right)R_{a,\beta}\left(x\right)$
and $\mathbf{J}_{aL}\left(x\right)=L_{a,\alpha}^{\dagger}\left(x\right)\left(\frac{\boldsymbol{\sigma}_{\alpha,\beta}}{2}\right)L_{a,\beta}\left(x\right)$
(with $a=1,2$) are the smooth components of the spin density, with
bosonic representation\begin{widetext}

\begin{align}
J_{a\lambda}^{x}\left(x\right) & =\frac{1}{2\pi a}\cos\left\{ \sqrt{2}\left[\lambda\phi_{a,s}\left(x\right)-\theta_{a,s}\left(x\right)\right]\right\} ,\label{eq:Jx}\\
J_{a\lambda}^{y}\left(x\right) & =\frac{1}{2\pi a}\sin\left\{ \sqrt{2}\left[\lambda\phi_{a,s}\left(x\right)-\theta_{a,s}\left(x\right)\right]\right\} ,\label{eq:Jy}\\
J_{a\lambda}^{z}\left(x\right) & =-\frac{1}{\sqrt{8}\pi}\left[\partial_{x}\phi_{a,s}\left(x\right)-\lambda\partial_{x}\theta_{a,s}\left(x\right)\right],\label{eq:Jz}
\end{align}
where $\lambda=R(L)$ corresponds to the plus (minus) sign, and where $\mathbf{N}_{a}\left(x\right)=R_{a,\alpha}^{\dagger}\left(x\right)\left(\frac{\boldsymbol{\sigma}_{\alpha,\beta}}{2}\right)L_{a,\beta}\left(x\right)+\mbox{H.c.}$,
are the staggered components

\begin{alignat}{1}
\mathbf{N}_{1}\left(x\right) & =\frac{\cos\left[\sqrt{2}\phi_{1c}\left(x\right)\right]}{\pi a}\left(\cos\left[\sqrt{2}\theta_{1s}\left(x\right)\right],-\sin\left[\sqrt{2}\theta_{1s}\left(x\right)\right],-\cos\left[\sqrt{2}\phi_{1s}\left(x\right)\right]\right),\label{eq:N1}\\
\mathbf{N}_{2}\left(x\right) & =\frac{m_{2}}{\pi a}\left(\cos\left[\sqrt{2}\theta_{2s}\left(x\right)\right],-\sin\left[\sqrt{2}\theta_{2s}\left(x\right)\right],-\cos\left[\sqrt{2}\phi_{2s}\left(x\right)\right]\right),\label{eq:N2}
\end{alignat}
\end{widetext}with $m_{2}=\left\langle \cos\left[\sqrt{2}\phi_{2c}\left(x\right)\right]\right\rangle $
resulting from the integration of the gapped charge degrees of freedom
in the Heisenberg chain. Therefore, although the spin densities (\ref{eq:pi_density})
and (\ref{eq:S_density}) look similar in the bosonized language,
they actually differ in two crucial aspects: 1) while in Eq. (\ref{eq:N2})
the charge degrees of freedom are absent in the expression of the
staggered magnetization, they are still present in (\ref{eq:N1})
in the term $\cos\left(\sqrt{2}\phi_{1c}\left(x\right)\right)$ and
we need to consider them. 2) Comparing Eqs. (\ref{eq:pi_density})
and (\ref{eq:S_density}), we note a sign difference in the staggered
components. This sign is related to the $p$-wave nature of the operators
$p_{j,\sigma}$, and is therefore intimately connected to the topology
of the Kondo interaction. The role of this sign turns out to be crucial
in the rest of the paper.

Replacing the above results into Eq. (\ref{eq:H_K}), and taking the
continuum limit, the Kondo interaction becomes in the bosonic language

\begin{align}
H_{K} & \sim2J_{K}a\int_{0}^{L_c}dx\biggl[\sum_{\lambda,\lambda^{\prime}=L,R}:\mathbf{J}_{1\lambda}\left(x\right)\cdot\mathbf{J}_{2\lambda^{\prime}}\left(x\right):\nonumber \\
 & -:\mathbf{N}_{1}\left(x\right)\cdot\mathbf{N}_{2}\left(x\right):\biggr],\label{eq:H_K_spin_densities}
\end{align}
where the sign in the second line is a consequence of the above mentioned
sign in the staggered part of $\boldsymbol{\pi}\left(x\right)$.

Note that this model is reminiscent of the (non-topological) 1D Kondo-Heisenberg
model, which has recently received much attention in the context of
pair-density wave ordered phases in high-$T_{c}$ cuprate physics \cite{zachar_kondo_chain_toulouse,Sikkema97_Spin_gap_in_a_doped_Kondo_chain,zachar_exotic_kondo,Zachar01_Staggered_phases_1D_Kondo_Heisenberg_model,Berg10_Pair_density_wave_in_Kondo_Heisenberg_Model,Dobry13_SC_phases_in_the_KH_model,Cho14_Topological_PDW_superconducting_state_in_1D},
and to the Hamiltonian of a spin-1/2 ladder  \cite{shelton_spin_ladders,gogolin_disordered_ladder,lecheminant02_magnet, Robinson12_Finite_wavevector_pairing_in_doped_ladders}.
However, a crucial difference with those works is the non-trivial
structure of the Kondo interaction, which differs from the usual coupling
$\sim J_{K}\mathbf{S}_{j}\cdot\mathbf{s}_{j}$, where $\mathbf{s}_{j}\equiv c_{j,\alpha}^{\dagger}\left(\frac{\boldsymbol{\sigma}_{\alpha\beta}}{2}\right)c_{j,\beta}$
is the standard (i.e., $s$-wave in this context) spin density in
the fermionic chain. The first line in (\ref{eq:H_K_spin_densities})
is in fact closely related to the model considered in Refs.  
\cite{Sikkema97_Spin_gap_in_a_doped_Kondo_chain,zachar_exotic_kondo,Zachar01_Staggered_phases_1D_Kondo_Heisenberg_model,Berg10_Pair_density_wave_in_Kondo_Heisenberg_Model,Dobry13_SC_phases_in_the_KH_model}.
In the half-filling situation we are analyzing here, however, the
most relevant part of $H_{K}$ (in the RG sense) is given by the product
$\mathbf{N}_{1}\left(x\right)\cdot\mathbf{N}_{2}\left(x\right)$,
which dominates the physics at low energies  \cite{shelton_spin_ladders,gogolin_disordered_ladder,lecheminant02_magnet, Robinson12_Finite_wavevector_pairing_in_doped_ladders}.
The term $\mathbf{N}_{1}\left(x\right)\cdot\mathbf{N}_{2}\left(x\right)$ only survives at half filling, and when both chains
have the same lattice parameter (in other situations, the oscillatory factors $~e^{\pm i2k_F x}$ suppress this term, and the situation corresponds to the case analyzed in Refs. \cite{Sikkema97_Spin_gap_in_a_doped_Kondo_chain,zachar_exotic_kondo,Zachar01_Staggered_phases_1D_Kondo_Heisenberg_model,Berg10_Pair_density_wave_in_Kondo_Heisenberg_Model,Dobry13_SC_phases_in_the_KH_model}). 
Therefore, for our present purposes,
we can neglect the first term in Eq. (\ref{eq:H_K_spin_densities})
and focus on the second term 
\begin{align}
H_{K} & \approx-2J_{K}a\int_{0}^{L_c}dx\;\mathbf{N}_{1}\left(x\right)\cdot\mathbf{N}_{2}\left(x\right),\nonumber \\
 & =-\frac{2J_{K}m_{2}}{\pi^{2}a}\int_{0}^{L_c}dx\;\cos\left(\sqrt{2}\phi_{1c}\right)\nonumber \\
 & \left[\cos\left(\sqrt{2}\theta_{1s}\right)\cos\left(\sqrt{2}\theta_{2s}\right)+\sin\left(\sqrt{2}\theta_{1s}\right)\right.\nonumber \\
 & \left.\times\sin\left(\sqrt{2}\theta_{2s}\right)+\sin\left(\sqrt{2}\phi_{1s}\right)\sin\left(\sqrt{2}\phi_{2s}\right)\right].\label{eq:H_K_bosonic}
\end{align}
At this point we note that the problem is reminiscent of the well-known
case of $S=1/2$ ladders with open boundary conditions  \cite{gogolin_disordered_ladder,lecheminant02_magnet},
with the important difference that here there is an extra factor $\sim\cos\left(\sqrt{2}\phi_{1c}\right)$.

The physics of the spin sector {[}i.e., term in square brackets in
(\ref{eq:H_K_bosonic}){]} is quite non-trivial due to the presence
of both the canonically conjugate fields $\phi_{a,s}\left(x\right)$
and $\theta_{a,s}\left(x\right)$, which cannot be simultaneously
stabilized. However, the analysis of the charge sector is simpler,
as only the field $\phi_{1c}\left(x\right)$ appears in the expression.
This means that in the limit $J_{K}\rightarrow\infty$ the system
can gain energy by ``freezing out'' the charge degrees of freedom,
i.e., $m_{1}=\left\langle \cos\left[\sqrt{2}\phi_{1c}\left(x\right)\right]\right\rangle $,
as there is no other competing mechanism. In the next Section we substantiate
these ideas by providing a rigorous analysis.

\section{\label{sec:RG}Renormalization-group analysis}

Based on the similarity with the physics of spin ladders, we introduce
symmetric and antisymmetric fields $\phi_{\pm}=\frac{1}{\sqrt{2}}\left(\phi_{1s}\pm\phi_{2s}\right)$
and $\theta_{\pm}=\frac{1}{\sqrt{2}}\left(\theta_{1s}\pm\theta_{2s}\right)$,
in terms of which the Hamiltonian becomes

\begin{align}
H_{K} & =-\frac{J_{K}m_{2}}{\pi^{2}a}\int_{0}^{L_c}dx\;\cos\left(\sqrt{2}\phi_{1c}\right)\nonumber \\
 & \times\left[-\cos\left(2\phi_{+}\right)+\cos\left(2\phi_{-}\right)+2\cos\left(2\theta_{-}\right)\right],\label{eq:H_K_bosonic_pm}
\end{align}
In what follows, we assume identical spinon dispersion $v_{1s}=v_{2s}=v_{s}$.
Although this assumption is certainly an idealization, one can show
that the asymmetry $\delta v\equiv v_{1s}-v_{2s}$ is an irrelevant
perturbation in the renormalization-group sense, and therefore we
do not expect that small asymmetries will have a qualitative effect
on our results. We now write the Euclidean action of the system using
complex space-time coordinates $z=v_{F}\tau+ix$ and $\overline{z}=v_{F}\tau-ix$,
with $\tau=it$ the imaginary time, and the left and right fields
$\phi_{\nu}=\left(\phi_{\nu L}+\phi_{\nu R}\right)/2$, where $\{\nu=+,-,1c\}$.
The Euclidean action becomes: 
\begin{align}
S & =S_{0}+S_{U}+S_{K},\label{S_total}\\
S_{0} & =-\frac{1}{4\pi}\int d^{2}r\Big\{\left(\partial_{z}\phi_{cL}\right)^{2}+\left(\partial_{\overline{z}}\phi_{cR}\right)^{2}\nonumber \\
+ & \sum_{\nu=\pm}\left[\left(\partial_{z}\phi_{\nu R}\right)^{2}+\left(\partial_{\overline{z}}\phi_{\nu R}\right)^{2}\right]\Big\},\\
S_{U} & =G_{2c}\int d^{2}r\; O_{2c}\left(r\right)+G_{3}\int d^{2}r\; O_{3}\left(r\right),\label{eq:Sfixpoint}\\
S_{K} & =G_{K}\int\frac{d^{2}r}{\sqrt{a}}\; O_{K}\left(r\right),
\end{align}
with $d^{2}r=v_{F}dxd\tau$, and where we have defined the dimensionless
couplings: 
\begin{align}
G_{2c}^{0}=G_{3}^{0}=\frac{Ua}{\pi v_{F}},\qquad & G_{K}^{0}=\frac{J_{K}m_{2}a}{\pi v_{F}}\label{eq:G^0}
\end{align}
and the scaling operators: 
\begin{align}
O_{2c} & =\frac{1}{4\pi}\partial_{z}\phi_{1cL}\partial_{\overline{z}}\phi_{1cR},\label{eq:O2c}\\
O_{3} & =-\frac{2\pi}{L^{2}}\colon\cos\left(2\sqrt{2}\phi_{1c}\right)\colon,\label{eq:O3}\\
O_{K} & =-\frac{\sqrt{2\pi}}{L^{3/2}}\colon\cos\left(\sqrt{2}\phi_{1c}\right)\left[-\cos\left(2\phi_{+}\right)\right.\nonumber \\
 & \left.+\cos\left(2\phi_{-}\right)+2\cos\left(2\theta_{-}\right)\right]\colon,\label{eq:OK}
\end{align}
where we have explicitly normal-ordered the operators. $S_{0}$ corresponds
to a free fixed-point action, and $S_{U}$ and $S_{K}$ are perturbations
arising from the Hubbard repulsion in chain $1$ and the Kondo inter-chain
interaction, respectively. We have neglected all the perturbations
in the spin sector generated by $U$, as they renormalize to zero
along the SU(2)-invariant line in the parameter space. We have also
neglected the less relevant terms coming from the product of the smooth
part of the currents in equations (\ref{eq:pi_density}) and (\ref{eq:S_density}).

Expanding the generating functional $Z=\int\prod_{i=\left\{ 1c,\pm\right\} }\mathcal{D}\left[\phi_{i},\theta_{i}\right]\; e^{-S\left[\phi_{i},\theta_{i}\right]}$
perturbatively at second order in $G_{i}$, we obtain the product
of the different operators (i.e., the operator product expansion or
OPE) of Eqs. (\ref{eq:O2c}), (\ref{eq:O3}) and (\ref{eq:OK}). Importantly,
the OPE of the Kondo interaction $O_{K}\left(z^{\prime},\overline{z}^{\prime}\right)O_{K}\left(z,\overline{z}\right)$
gives rise to operators $O_{3}$ and $O_{2c}$, which are already
present in the charge sector of chain $1$. This corresponds to an
effective dynamically generated Hubbard repulsion originated by the
inter-chain Kondo coupling (see Appendix \ref{sec:appendix_RG}).
Therefore, even for an initially non-interacting chain (i.e., $U=0$),
this emergent repulsive interaction induces the opening of a Mott
insulating gap in the charge sector of the half-filled conduction
band. At energies below this gap, the field $\phi_{1c}$ becomes pinned
to the degenerate values $0$ or $\frac{\pi}{\sqrt{2}}$. Note that
only the latter is consistent with the boundary conditions given in
Eq. (\ref{eq:constraint_phi1c}). Therefore, this analysis suggests
that the energy is minimized by a uniform configuration of the field
$\phi_{1c}$, which ``freezes'' at the bulk value $\frac{\pi}{\sqrt{2}}$. While other configurations with kink excitations connecting the different minima are certainly possible, these configurations are more costly energetically speaking and do not belong to the groundstate. Therefore, energy minimization prevents the ground state from developing
``kink'' excitations in the charge sector and, consequently, we can exclude the presence
of localized charge edge-states. Then, at low enough energies, the
system becomes a Mott insulator in the bulk due the electronic correlations,
and no topological effects arise in the charge sector.

A more quantitative study can be done by analyzing the two-loop RG-flow
equations (see Appendix \ref{sec:appendix_RG} for details): 
\begin{align}
\frac{dG_{2c}}{dl} & =G_{3}^{2}+\frac{3}{4}G_{K}^{2},\label{eq:RGG2c}\\
\frac{dG_{3}}{dl} & =G_{2c}G_{3}+\frac{3}{4}G_{K}^{2},\label{eq:RGG3}\\
\frac{dG_{K}}{dl} & =\frac{G_{K}}{2}+\frac{1}{4}G_{2c}G_{K}+G_{3}G_{K},\label{eq:RGGK}
\end{align}
where $l=\ln\left(a/a_{0}\right)$ is the logarithmic RG scale. When
$G_{K}=0$ these equations reduce to the Kosterlitz-Thouless ones
corresponding to the charge sector of the Hubbard model. They predict
a charge gap which is exponentially small in $U$  \cite{giamarchi_book_1d}.

Let us now analyze the case of an initially $U=0$. In this situation,
only the linear term survives in the right hand side of Eq. (\ref{eq:RGGK}),
which expresses the relevance of the operator $O_{K}$, and gives rise
to an exponential increase of $G_{K}\left(l\right)$ with the RG scale
$l$ as $G_{K}\left(l\right)=G_{K}^{0}e^{\frac{l}{2}}$. As a consequence,
the coupling $G_{3}\left(l\right)$ in Eq. (\ref{eq:RGG3}), representing
the Hubbard repulsion, also increases exponentially $G_{3}\left(l\right)=\frac{3}{4}\left(G_{K}^{0}\right)^{2}\left(e^{l}-1\right)$,
from its initial value $G_{3}\left(0\right)=0$. This produces the
anticipated Mott insulating gap $\Delta_{c}$ in the charge sector.
Its dependence with the parameters can be obtained by the procedure
described in page 65 of Ref. \cite{giamarchi_book_1d}. We obtain
$\Delta_{c}\sim\left(G_{K}^{0}\right)^{2}$ for small enough $J_{K}$.
We could envisage that if $U$ and $J_{K}$ would be of the same order,
the dominant contribution to $\Delta_{c}$ would come from the inter-chain
Kondo coupling.

Following previous references  \cite{shelton_spin_ladders,gogolin_disordered_ladder,lecheminant02_magnet,Robinson12_Finite_wavevector_pairing_in_doped_ladders},
we can refermionize Eq. (\ref{eq:H_K_bosonic_pm}) noting that the
scaling dimension of the cosines in the square bracket exactly corresponds
to the free-fermion point and therefore they can be written in terms
of right and left-moving free Dirac fermion fields $\eta_{\pm,R}\left(x\right)$
and $\eta_{\pm,L}\left(x\right)$ as

\begin{align}
\cos\left(2\phi_{\pm}\right) & =-i\pi a\left(\eta_{\pm R}^{\dagger}\eta_{\pm L}-\eta_{\pm L}^{\dagger}\eta_{\pm R}\right),\label{eq:refermionization_phi_pm}\\
\cos\left(2\theta_{\pm}\right) & =i\pi a\left(\eta_{\pm R}^{\dagger}\eta_{\pm L}^{\dagger}-\eta_{\pm L}\eta_{\pm R}\right).\label{eq:refermionization_theta_pm}
\end{align}
For later purposes, it is more convenient to introduce a Majorana-fermion
representation of the fields $\eta_{+,\lambda}=\frac{1}{\sqrt{2}}\left(\xi_{\lambda}^{2}+i\xi_{\lambda}^{1}\right)$,
$\eta_{-,\lambda}=\frac{1}{\sqrt{2}}\left(\xi_{\lambda}^{3}+i\xi_{\lambda}^{0}\right)$,
the Hamiltonian can be compactly written as $H=H_{0}+H_{K}$, where

\begin{align}
H_{0} & =\frac{v_{1c}}{2\pi}\int_{0}^{L_c}dx\left[\frac{\left(\partial_{x}\phi_{1c}\right)^{2}}{K_{1c}}+K_{1c}\left(\partial_{x}\theta_{1c}\right)^{2}-\frac{U\cos\sqrt{8}\phi_{1c}}{v_{1c}\pi a^{2}}\right]\nonumber \\
 & -i\frac{v_{s}}{2}\sum_{a=0}^{3}\int_{0}^{L_c}dx\left[\xi_{R}^{a}\partial_{x}\xi_{R}^{a}-\xi_{L}^{a}\partial_{x}\xi_{L}^{a}\right],\label{eq:H0_boson_fermion}\\
H_{K} & =i\frac{J_{K}m_{2}}{2\pi}\int_{0}^{L_c}dx\;\cos\left(\sqrt{2}\phi_{1c}\right)\left[3\xi_{R}^{0}\xi_{L}^{0}-\sum_{a=1}^{3}\xi_{R}^{a}\xi_{L}^{a}\right],\label{eq:HK_boson_fermion}
\end{align}
where the Majorana fields obey the boundary conditions

\begin{align}
\xi_{R}^{a}\left(0\right) & =\xi_{L}^{a}\left(0\right),\label{eq:constraint_xi_0}\\
\xi_{R}^{a}\left(L_c\right) & =\xi_{L}^{a}\left(L_c\right).\label{eq:constraint_xi_L}
\end{align}
The uniform symmetric and antisymmetric spin densities in the ladder
become \cite{shelton_spin_ladders}

\begin{align}
M_{\lambda}^{a} & =J_{1,\lambda}^{a}+J_{2,\lambda}^{a}=-\frac{i}{2}\epsilon_{abc}\xi_{\lambda}^{b}\xi_{\lambda}^{c},\label{eq:M}\\
K_{\lambda}^{a} & =J_{1,\lambda}^{a}-J_{2,\lambda}^{a}=-\frac{i}{2}\xi_{\lambda}^{0}\xi_{\lambda}^{a},\label{eq:K}
\end{align}
with $a=\left\{ 1,2,3\right\} $ and $\lambda=\left\{ R,L\right\} $.
This is a well-known representation of two independent $\text{SU}\left(2\right)_{1}$
Kac-Moody currents $\mathbf{J}_{1,\lambda}$ and $\mathbf{J}_{2,\lambda}$
in terms of four Majorana fields \cite{shelton_spin_ladders}. In our case, these four degrees
of freedom, resulting from the combination of the two SU(2) spin density
fields in the two chains, are expressed in terms of a singlet $\xi_{\lambda}^{0}$
and triplet $\xi_{\lambda}^{a}$ Majorana fields.

From the previous analysis we conclude that at temperatures $T\ll\Delta_{c}$,
the charge and spin degrees of freedom become effectively decoupled,
and the low-energy Hamiltonian of the system can be written as $H\rightarrow\tilde{H}=\tilde{H}_{c}+\tilde{H}_{s}$,
with: 
\begin{align}
\tilde{H}_{c}=\frac{v_{1c}}{2\pi}\int_{0}^{L_c}\!\! dx\!\left[\frac{\left(\partial_{x}\phi_{1c}\right)^{2}}{K_{1c}}+K_{1c}\left(\partial_{x}\theta_{1c}\right)^{2}+\frac{U_\text{eff}m_{1}^{2}}{v_{1c}\pi a^{2}}\left(\phi_{1c}\right)^{2}\right],\label{eq:H_tilde_c}
\end{align}
where, based on the discussion above Eq. (\ref{eq:RGG2c}), we have expanded the charge field near the value $\phi_{1c}=\pi/\sqrt{2}$, and
\begin{align}
\tilde{H}_{s}= & -i\frac{v_{s}}{2}\sum_{a=0}^{3}\int_{0}^{L_c}dx\left[\xi_{R}^{a}\partial_{x}\xi_{R}^{a}-\xi_{L}^{a}\partial_{x}\xi_{L}^{a}\right]\nonumber \\
 & +i\frac{J_{K}m_{2}m_{1}}{2\pi}\int_{0}^{L_c}dx\;\left[3\xi_{R}^{0}\xi_{L}^{0}-\sum_{a=1}^{3}\xi_{R}^{a}\xi_{L}^{a}\right],\label{eq:H_tilde_s}
\end{align}
where $U_\text{eff} \equiv U\left(J_K\right)$ is the renormalized Coulomb repulsion parameter and 
where the charge degrees of freedom in the fermionic chain develop
a gap $m_{1}\equiv m_{1}\left(J_{K}\right)=\langle\cos\left(\sqrt{2}\phi_{1c}\right)\rangle$.
Note that, in this form, the spin Hamiltonian $\tilde{H}_{s}$ is
similar to that of a spin-1/2 ladder  \cite{shelton_spin_ladders,gogolin_disordered_ladder,lecheminant02_magnet}.
The quantitative determination of the parameter $m_{1}$ requires
a self-consistent calculation, which is beyond the scope of the present
work. Nevertheless, the previous analysis allows to understand two
important aspects of the topological Kondo-Heisenberg chain: a) the
generation of an insulating state in the bulk (necessary to reproduce
the bulk insulating state of a TKI), and b) the emergence of magnetic
edge-states, which is the subject of the next section.

\section{\label{sec:topo}$S=1/2$ magnetic edge-modes and topological invariant}

\begin{figure}
\includegraphics[bb=60bp 0bp 560bp 380bp,clip,scale=0.45]{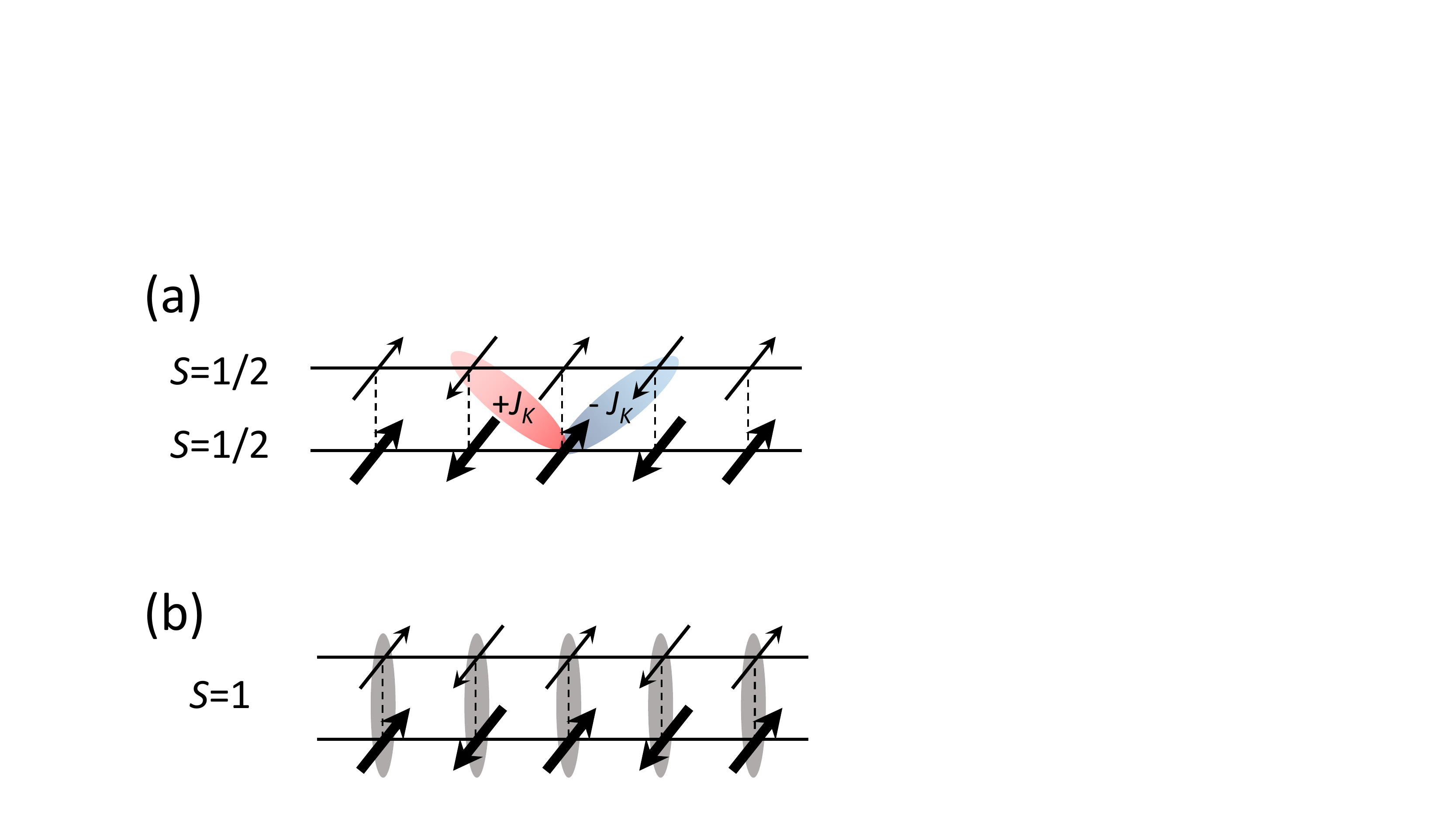}\protect\caption{\label{ladder}(a) At temperatures $T\ll\Delta_{c}\sim J_{K}^{2}$,
the dynamically induced Mott gap, the charge degrees of freedom in
the Hubbard chain become frozen and the system maps onto a ladder
where the non-local antiferromagnetic Kondo coupling effectively induces
\textit{local ferromagnetic} correlations between the spins in the
ladder (vertical dashed lines). (b) The system depicted in (a) can
be mapped onto a $S=1$ Haldane chain, which supports topologically
protected $S=1/2$ end-states  \cite{shelton_spin_ladders,gogolin_disordered_ladder,lecheminant02_magnet,affleck_klt_short,affleck_klt_long,ng_schwinger,ng_berry,ng_edges}.}
\end{figure}

The previous analysis shows that, at low temperatures $T\ll\Delta_{c}$,
the 1D ``$p-$wave'' Kondo-Heisenberg chain at half-filling can
be effectively mapped onto a spin ladder problem, which is dominated
by the staggered components of the spin densities. Interestingly,
for an antiferromagnetic Kondo coupling $J_{K}>0$, the negative sign
emerging from the structure of the non-local Kondo interaction (\ref{eq:H_K_bosonic}),
effectively induces a\textit{local ferromagnetic interaction} {[}see
vertical dashed lines in Fig. \ref{ladder}(a){]}. It is well-known
that the spin ladder with ferromagnetic exchange coupling $J_{\perp}<0$
along the rungs has a low-energy triplet sector\cite{shelton_spin_ladders,gogolin_disordered_ladder,lecheminant02_magnet}, which maps onto the Haldane spin-1 chain \cite{haldane_ising,haldane_gap}.
Therefore, at temperatures $T\ll J_{\perp}$, our model describes
the physics of the Haldane spin-1 chain,
which is known to host symmetry-protected topological spin-1/2 modes
at the boundaries \cite{affleck_klt_short,affleck_klt_long,ng_schwinger,ng_berry,ng_edges,Berg08_Hidden_string_order_in_lattice_bosons,Gu09_RG_for_tensor_networds_and_SPT_order,Pollmann10_Entaglement_spectrum_of_topological_phases_in_1D,Pollmann12_SPT_phases_in_1D}.
This situation is also very reminiscent to the case of the ferromagnetic
Kondo lattice \cite{Tsunetsugu92_half_filled_Kondo_lattice_1D,Garcia04_Spin_order_in_1D_Kondo_and_Hund_lattices,Janani14_Hubbard_model_on_triangular_necklace,Janani14_Topological_spin_liquid_in_1D}.
To see how these spin-1/2 boundary modes emerge in our low-energy Hamiltonian
(\ref{eq:H_tilde_s}), we consider solutions of the eigenvalue equation
$\tilde{H}_{s}\Psi^{a}\left(x\right)=0$, where $\Psi^{a}\left(x\right)=\left(\xi_{R}^{a},\xi_{L}^{a}\right)^{T}$
is a Majorana spinor, and $a=\left(0,1,2,3\right)$  \cite{lecheminant02_magnet}.
In matrix form, this equation is

\begin{align}
\left[-iv_{s}\hat{\tau}_{3}\partial_{x}+m_{a}\hat{\tau}_{2}\right]\Psi^{a}\left(x\right) & =0,\label{eq:eq_zero_mode}
\end{align}
with $m_{a}=-3J_{K}m_{2}m_{1}/2\pi$ for $a=0$ {[}$m_{a}=J_{K}m_{2}m_{1}/2\pi$
for $a=\left(1,2,3\right)${]} and $\hat{\tau}_{i}$ the Pauli matrices
acting of the right-, left-moving space. Eq. (\ref{eq:eq_zero_mode})
admits solutions of the form%
{} $\Psi^{a}\left(x\right)\propto\exp\left(-m_{a}\hat{\tau}_{1}x/v_{s}\right)\Psi^{a}\left(0\right)$
and one would think that in principle two normalizable solutions in
the limit $x\rightarrow\infty$ could arise: 1) the choice $\Psi^{a}\left(0\right)=\left(1,-1\right)^{T}$
with $m_{a}<0$, and 2) $\Psi^{a}\left(0\right)=\left(1,1\right)^{T}$
with $m_{a}>0$. However, note that only the last choice is compatible
with the boundary condition (\ref{eq:constraint_xi_0}). Then, the
only physical solution localized around $x=0$ corresponds to the
choice $m_{a}>0$, which in our case corresponds to $a=\left(1,2,3\right)$

\begin{align}
\Psi^{a}\left(x\right)\propto\xi_{a}e^{-m_{a}x/v_{s}} & \left(\begin{array}{c}
1\\
1
\end{array}\right)\quad a=\left(1,2,3\right),\label{eq:end_Majorana}
\end{align}
with $\xi_{a}$ being a localized Majorana fermion. Using the expression
(\ref{eq:M}) for the smooth part of the spin density, we can physically
associate the presence of the localized Majorana bound-state with
a localized spin-1/2 magnetic edge-state. This is consistent with
the results in Ref. \cite{Alexandrov2014_End_states_in_1DTKI} where,
in the case of a uniform Kondo interaction, a magnetic mode with no
admixture with charge degrees of freedom emerges at the boundaries.
However, note that the origin of these edge-states is quite different:
while in the mean-field regime they emerge as a consequence of Kondo-unscreened
end-spins in the Heisenberg chain, in our case they are intimately
related to the physics of the Haldane chain.

We now derive a topological invariant to characterize the presence
of the edge-states, using a suitable generalization of the concept of electrical polarization in 1D insulators  \cite{Resta99_Electron_localization, Ortiz94_Many_body_formulation_of_polarization, Aligia05_Dimerized_ionic_Hubbard, Torio06_Generalized_ionic_Hubbard, Xiao10_Berry_phase_in_electronic_properties,Fu'06}. We therefore focus on the uniform magnetization
Eq. (\ref{eq:M}). Although the original Hamiltonian has spin-rotational
symmetry, the Abelian bosonization is not an explicitly SU(2)-invariant
formalism. Therefore, while the choice of axes is arbitrary due to
the spin rotation symmetry of the problem, once we define the $z-$direction
as the spin-quantization axis, the perpendicular components of the
magnetization $M^{x}\left(x\right)$ and $M^{y}\left(x\right)$ acquire
a more complicated mathematical form. For that reason, it is 
convenient to focus only on the $z$ component of the symmetric spin
density Eq. (\ref{eq:M})
\begin{align}
M^{z}\left(x\right) & =\sum_{\lambda=L,R}M_{1,\lambda}^{z}\left(x\right)+M_{2,\lambda}^{z}\left(x\right),\nonumber \\
 & =-\frac{1}{\sqrt{\pi}}\partial_{x}\phi_{+}\left(x\right).\label{eq:Mz_total_bosonic}
\end{align}
In the expression above, we have used Eq. (\ref{eq:Jz}) and the
definition of $\phi_{+}\left(x\right)=\frac{1}{\sqrt{2}}\left[\phi_{1}\left(x\right)+\phi_{2}\left(x\right)\right]$.
We now define the \textit{total magnetic moment} along the $\hat{z}-$axis at one end (for concreteness,
the left end) of the chain as $m_{T}^{z}\equiv \frac{1}{\sqrt{\pi}}\int_{0}^{x_{b}}dx\; M^{z}\left(x\right)$,
where $x_{b}$ is an unspecified position in the interior of the chain
where the magnetization reaches the value in the bulk. In bosonic
language, it acquires the compact form

\begin{align}
m_{T}^{z} & =-\frac{1}{\pi}\left[\phi_{+}\left(x_{b}\right)-\phi_{+}\left(0\right)\right].\label{eq:Mz_total_bosonic_2}
\end{align}
From the expression of the bosonic Hamiltonian Eq. (\ref{eq:H_K_bosonic_pm})
in the limit $J_{K}\rightarrow\infty$, we see that the system minimizes
the energy in the bulk by pinning the field $\phi_{+}\left(x\right)$
to one of the degenerate values

\begin{align}
\phi_{+}\left(x_{b}\right) & =\pm\frac{\pi}{2}.\label{eq:phi_plus_bulk}
\end{align}
On the other hand, from the definition of $\phi_{\pm}\left(x\right)=\frac{1}{\sqrt{2}}\left[\phi_{1s}\left(x\right)\pm\phi_{2s}\left(x\right)\right]$
and Eq. (\ref{eq:constraint_phi1s}), the boundary condition $\phi_{+}\left(0\right)=0$
is obtained. Replacing these values into Eq. (\ref{eq:Mz_total_bosonic_2}),
we obtain the following quantized values of the magnetic moment at the
left end
\begin{align}
m_{T}^{z} & =\pm\frac{1}{2}.\label{eq:Mz_total_bosonic_3}
\end{align}
This magnetic moment at the end of the chain is analogous to the electrical
polarization \cite{Resta99_Electron_localization, Ortiz94_Many_body_formulation_of_polarization, Aligia05_Dimerized_ionic_Hubbard, Torio06_Generalized_ionic_Hubbard, Xiao10_Berry_phase_in_electronic_properties}
or the time-reversal polarization \cite{Fu'06} in 1D insulators. In particular, we note the close relation between our formula Eq. \ref{eq:Mz_total_bosonic_2} and the expressions for the displacement operator appearing in Eq. (23) of Ref. \cite{Aligia05_Dimerized_ionic_Hubbard}, and for the time-reversal polarization appearing in Eq. (4.8) in Ref. \cite{Fu'06}, both given in bosonic language. Frome here, we can define a $Z_{2}-$topological invariant which characterizes
the topological phase of the Kondo-Heisenberg chain
\begin{align}
Q & =\left(-1\right)^{2m_{T}^{z}/\pi}=e^{i2\pi m_{T}^{z}},\label{eq:topological_invariant}
\end{align}
which in the limit of an infinite system $L\rightarrow\infty$ is
$Q=-1$ in the topological phase ($J_{K}>0$), and $Q=1$ in the trivial
one ($J_{K}<0$).

The bosonic representation also provides an alternative way to demonstrate
the existence of magnetic edge modes. Since none of the degenerate
values (\ref{eq:phi_plus_bulk}) of $\phi_{+}\left(x\right)$ in the
bulk satisfy the boundary condition at $x=0$, we conclude that a
kink excitation necessarily must emerge near the boundary in order
to connect those values: precisely this kink excitation gives rise
to the spin-1/2 end-state, upon use of Eq. (\ref{eq:Mz_total_bosonic}).
We remind the reader that in Section \ref{sec:RG}, using similar
arguments, we demonstrated the absence of kink configurations in the
charge field $\phi_{1c}\left(x\right)$, and the fact that there are
no charge edge-states in the ground state.

\section{\label{sec:conclusions}Conclusions}

We have studied theoretically a model for a topological 1D Kondo insulator
(the 1D Kondo-Heisenberg
model coupled in the $p$-wave channel, with an on-site Hubbard
interaction $U$ in the conduction band) using the Abelian bosonization formalism, and derived the two-loop
RG flow equations for the system at half-filling. Our RG analysis
shows that the system develops a Mott-insulating gap at low enough
temperatures, even if $U=0$. Moreover, the remaining spin degrees
of freedom are effectively described by a ferromagnetic spin-1/2 ladder,
which in turn maps onto a spin-1 Haldane chain with topologically
protected spin-1/2 magnetic edge-modes. This situation is reminiscent
to the physics of the ferromagnetic Kondo necklace, which also maps
onto the spin-1 Haldane chain \cite{Tsunetsugu92_half_filled_Kondo_lattice_1D,Garcia04_Spin_order_in_1D_Kondo_and_Hund_lattices,Janani14_Hubbard_model_on_triangular_necklace,Janani14_Topological_spin_liquid_in_1D},
although in our case it arises as a result of the non-trivial structure
of the Kondo coupling.

In contrast to three-dimensional bulk topological Kondo insulators,
where the mean-field approximation is well justified and the system
can be effectively described in terms of non-interacting quasiparticles
opening a (renormalized) hybridization gap near the Fermi surface
 \cite{Coleman84,coleman87,Dzero10_Topological_Kondo_Insulators},
in one spatial dimension the presence of strong quantum fluctuations
cannot be ignored, and one is forced to use different approaches. The Abelian
bosonization method allows to obtain a description of the 1D TKI which
is fundamentally different from the mean-field picture. In the first
place, the system develops a Mott gap (instead of a hybridization
gap) in the spectrum of charge excitations when the conduction band
is half-filled (small deviations from half-filling do not affect this
scenario qualitatively  \cite{giamarchi_book_1d}). This Mott gap
arises from umklapp processes at second order in the Kondo interaction.
Physically, this can be understood as a dynamically-induced effective
interaction term, which appears at order $J_{K}^{2}$ in the conduction
band by integrating out perturbatively short-time spin excitations
in the Heisenberg chain. In contrast to the mean-field description,
where the hybridization gap depends exponentially on the microscopic
Kondo coupling $\Delta_{c}\sim\exp\left(-1/J_{K}\right)$, the integration
of the RG Eqs. (\ref{eq:RGG2c})-(\ref{eq:RGGK}) in the limit $J_{K}\rightarrow0$
results in $\Delta_{c}\sim J_{K}^{2}$ . Our RG analysis indicates
that $J_{K}$ is a relevant perturbation and flows to strong coupling,
dominating the physics at low temperatures. In particular, at temperatures
below the Mott gap, the charge degrees of freedom are frozen and system
effectively behaves as a ferromagnetic spin-1/2 ladder, which is known
to map onto the spin-1 Haldane chain. Therefore, our work allows to
make an insightful connection between two a priori unrelated physical
models. Interestingly, exploiting this connection, we predict the
existence of topologically protected spin-1/2 edge states. This seems
to correspond to the ``magnetic phase'' found by Alexandrov and
Coleman \cite{Alexandrov2014_End_states_in_1DTKI}, which for a uniform
Kondo coupling $J_{K}$, is characterized by Kondo-unscreened spins
at the end of the Heisenberg chain. However, the emergence of these
edge states, again corresponds to a very different mechanism than
the one provided by the mean-field theory. Interestingly, within the
bosonization framework, we have been able to obtain a $Z_{2}$ topological
invariant {[}see Eq. (\ref{eq:topological_invariant}){]} in terms
of the magnetization at an end of the chain.

Our work opens the possibility to explore the physics of broken $Z_{2}\times Z_{2}$
hidden symmetry and the existence of a non-vanishing string order parameter
$\mathcal{O}_{\text{string}}^{\alpha}\equiv\lim_{\left|l-m\right|\rightarrow\infty}\left\langle S_{l}^{\alpha}e^{i\pi\sum_{l\leq j<m}S_{j}^{\alpha}}S_{m}^{\alpha}\right\rangle \neq0$
(which are well-known features of the Haldane phase  \cite{kennedy_z2z2_haldane,affleck_klt_short,affleck_klt_long})
in the $p-$wave Kondo-Heisenberg model. In particular, note the close
relation between the $Z_{2}$ topological invariant (\ref{eq:topological_invariant})
and the string-order parameter in bosonized form {[}see Eq. (83) in
Ref. \cite{shelton_spin_ladders}{]}.

Furthermore, we reiterate that the model studied here can be viewed
as a non-trivial strongly-correlated generalization of the old Tamm-Shockley
model \cite{Tamm32_Tamm_model,Shockley39_Shockley_model}. The latter
is a prototypical one-dimensional model that exhibits a topological
phase transition and it can be used to construct high-dimensional
topological band insulators \cite{Pershoguba12_Surface_states}.
Likewise, the strongly-correlated topological Kondo-Heisenberg model
could potentially become a building block in constructing higher-dimensional
strongly-interacting topological states -- not adiabatically connected
to ``simple'' topological band insulators. Although the physical realization of the 1D $p$-wave Kondo lattice model studied here in solid-state systems might be quite challenging, our results might have direct application to ultracold atom experiments, where double-well optical superlattices loaded with atoms in $s$ and $p$ orbitals have been realized \cite{Wirth11_Superfluidity_in_p_band_optical_lattice, Soltan-Panahi11_QPT_in_multiorbital_optical_lattices}. In such systems, one can imagine the atoms forming ladders where one of the legs corresponds to the $s$ orbitals and the other to $p$ orbitals (e.g., see Ref. \cite{Li13_Topological_interacting_ladders}). The overlap between $s$ and $p$ orbitals along the rungs vanish by symmetry, and therefore only the off-diagonal hopping $t_{sp}$ survives. Next, allowing for an on-site repulsive Hubbard $U^\prime$ interaction in the $s$-orbital leg (using, e.g., Feschbach resonances), one can derive an effective $p$-wave Kondo lattice model in the limit $t_{sp}/U \rightarrow 0$, introducing a canonical transformation to eliminate processes at first order in $t_{sp}$. At half-filling, the $s$-orbitals are effectively described by SU(2) spins and the Kondo parameter in our Eq. (\ref{eq:H_K}) becomes proportional to $J_K \sim t_{sp}^2/U$. Therefore,  the system can be described by the model described in this work.
Finally, we mention in
this context recent experimental results \cite{Weird_TKI_exp1,Weird_TKI_exp2,Weird_TKI_exp3},
which suggest the existence of a magnetic phase transition and/or
suppressed surface charge transport in select samples of Samarium
hexaboride ($\text{SmB}_{6}$ -- a three-dimensional topological Kondo
insulator). It is possible that these phenomena, which remain unexplained
at this stage, involve in a crucial way an interplay between band
topology and strong correlations, which conceivably may lead to the
formation of non-trivial magnetic topological surface modes reminiscent
to the edge states found here. In a more general context, our results
might be relevant to other materials which belong to the same ``Haldane
universality class'', thanks to the connection (unveiled in this
work) to the ferromagnetic Kondo lattice model. For example, the organic
molecular compound Mo$_{3}$S$_{7}$(dmit)$_{3}$ at two-third filling,
a promising candidate for a quantum spin liquid, has recently been
shown to be a realization of the ferromagnetic Kondo lattice model
at half-filling \cite{Janani14_Hubbard_model_on_triangular_necklace,Janani14_Topological_spin_liquid_in_1D},
and therefore it should realize a Haldane phase with magnetic end-modes
at low temperatures. 

A. M. L. acknowledges
support from National Science Foundation-Joint Quantum Institute-
Physics Frontier Center (NSF-JQ-PFC) and from program Red de Argentinos
Investigadores y Cient{\'i}ficos en el Exterior (RAICES), Argentina.
A. O. D. is partially supported by Proyecto de Investigaci{\'o}n Plurianual
(PIP) 11220090100392 of Consejo de Investigaciones Cient{\'i}ficas y T{\'e}cnicas
(CONICET), Argentina. V. G. was supported by Department of Energy-
Basic Energy Sciences (DOE-BES) DESC0001911 and Simons Foundation.
The authors would like to thank Thierry Giamarchi, Eduardo Fradkin,
Philippe Lecheminant, Edmond Orignac, Dmitry Efimkin, Xiaopeng Li
and Tigran Sedrakyan for useful discussions and for pointing out relevant
references.
\bibliographystyle{my_apsrev4-1}
%

\appendix
\numberwithin{equation}{section}

\section{\label{sec:appendix_RG}Dynamically-generated interactions and derivation
of the renormalization group equations}

In this Appendix we derive an effective action for the system and
the renormalization-group (RG) equations (\ref{eq:RGG2c}-\ref{eq:RGGK}).
The idea is to show that umklapp processes, which mimic a repulsive
interaction among electrons in the half-filled conduction band, arise
at order $\mathcal{O}\left(J_{K}^{2}\right)$ and open a gap in the
charge sector of the model. To that end, we expand the generating
functional of the system (i.e., the partition function) up to second
order in the coupling constants $G_{\alpha}$ following Refs. \cite{cardy_book_renormalization,fradkin_book}
\begin{widetext}

\begin{align}
\frac{Z}{Z_{0}} & =\left[1-\sum_{\alpha}\frac{G_{\alpha}}{a^{2-\Delta_{\alpha}}}\int d^{2}r\left\langle O_{\alpha}\right\rangle _{0}+\frac{1}{2}\sum_{\alpha,\beta}\frac{G_{\alpha}G_{\beta}}{a^{4-\Delta_{\alpha}-\Delta_{\beta}}}\int\int_{\left|r-r^{\prime}\right|>a}d^{2}rd^{2}r^{\prime}\left\langle O_{\alpha}\left(r\right)O_{\beta}\left(r^{\prime}\right)\right\rangle _{0}+...\right]\label{eq:Zperturb}
\end{align}
indexes $\alpha$ and $\beta$ run on $2c$, $3$ and $K$. Here,
$\Delta_{\alpha}$ is the scaling dimension of the operator $O_{\alpha}$
defined in Eqs. (\ref{eq:O2c}-\ref{eq:OK}) ($\Delta_{3}=\Delta_{2c}=2$,
$\Delta_{K}=\frac{3}{2}$), $Z_{0}=\int\prod_{i=\left\{ 1c,\pm\right\} }\mathcal{D}\left[\phi_{i},\theta_{i}\right]\; e^{-S_{0}\left[\phi_{i},\theta_{i}\right]}$
is the generating function of the free theory, and the mean values
$\left\langle \dots\right\rangle _{0}$ correspond also to that theory.
This formalism is standard in the analysis of 1D quantum systems,
and has been applied in several previous works (see for example Ref.
\cite{HirschModel_Dobry_Aligia} where the method is explained in
detail).

The third term of the r.h.s. in Eq. (\ref{eq:Zperturb}) takes the
same form as the second one if we assume that for $r\rightarrow r^{\prime}$
the product of two operators fulfills the following operator product
expansion (OPE) property \cite{cardy_book_renormalization,fradkin_book}
: 
\begin{align}
O_{\alpha}\left(r\right)O_{\beta}\left(r^{\prime}\right) & =\sum_{\gamma}C_{\alpha\beta}^{\gamma}\frac{O_{\gamma}\left(\frac{r+r^{\prime}}{2}\right)}{\mid r-r^{\prime}\mid^{\Delta_{\alpha}+\Delta_{\beta}-\Delta_{\gamma}}}+\mbox{more irrelevant operators.}\label{eq:defopes}
\end{align}
where $O_{\gamma}$ includes all the operators generated from each
OPE.

Let us focus on the OPE between two $O_{K}$ operators, which is the
most relevant perturbation in the RG sense, and is precisely the contribution
that leads to the umklapp processes we are trying to describe. To
simplify the discussion, here we return to the representation of the
bosonic fields in terms of left and right movers $\phi_{L}\left(z\right)$
and $\phi_{R}\left(\bar{z}\right)$ {[}see Eq. (\ref{eq:bosonizationRL}){]},
with $z=v_{F}\tau+ix$ and $\bar{z}=v_{F}\tau-ix$. We now assume
to be sufficiently deep in the bulk of the 1D system and far away
from the boundaries. In these conditions, the boundary conditions
(\ref{eq:boundary_condition_0}) and (\ref{eq:boundary_condition_L}),
and the commutation relation (\ref{eq:RL_commutation_relation}) can
be effectively neglected, and the fields $\phi_{L}\left(z\right)$
and $\phi_{R}\left(\bar{z}\right)$ become independent (i.e., they
do not mix). This allows us to focus only on the processes involving
the left-moving field $\phi_{L}\left(z\right)$(for right-moving fields
we just need to change $L\rightarrow R$ and $z\rightarrow\bar{z}$)
. The basic OPE we need is 
\begin{align}
\colon e^{i\lambda\phi_{L}\left(z\right)}\colon\colon e^{i\lambda^{\prime}\phi_{L}\left(z^{\prime}\right)}\colon & =\left(\frac{2\pi}{L}\right)^{\frac{\lambda\lambda^{\prime}}{2}}\colon e^{i\left(\lambda+\lambda^{\prime}\right)\phi_{L}\left(z^{\prime}\right)}\left[\frac{1}{\left(z-z^{\prime}\right)^{-\lambda\lambda^{\prime}}}+\frac{i\lambda\partial_{z^{\prime}}\phi_{L}\left(z^{\prime}\right)}{\left(z-z^{\prime}\right)^{-\lambda\lambda^{\prime}-1}}+...\right]\colon
\end{align}
\end{widetext}which was obtained by normal-ordering the rhs expression
and then developing for $z^{\prime}$ near $z$. Through repeated
use of this expression we obtain the desired OPE which reads:

\begin{align}
O_{K}\left(z,\overline{z}\right)O_{K}\left(z^{\prime},\overline{z}^{\prime}\right)= & -\frac{3}{4\pi}\frac{O_{3}}{\mid z-z^{\prime}\mid}-\frac{3}{4\pi}\frac{O_{2c}}{\mid z-z^{\prime}\mid}\nonumber \\
 & +\frac{3}{4\pi}\frac{O_{-}}{\mid z-z^{\prime}\mid}-\frac{1}{4\pi}\frac{O_{+}}{\mid z-z^{\prime}\mid},\label{eq:opeOKOK}
\end{align}
where we have defined the operators $O_{\pm}\equiv\frac{1}{4\pi}\partial_{\overline{z}}\phi_{\pm R}\partial_{z}\phi_{\pm L}$
{[}which also appear in the $z$-component of the product of the right
and left smooth-varying spin currents, in the first two lines in Eq.
(3.8) of Ref. \cite{Dobry13_SC_phases_in_the_KH_model}{]}. Note that
these terms break the SU(2) invariance of the model. This is a well-known
feature of the Abelian bosonization prescription, which is not explicitly
SU(2)-invariant formalism \cite{giamarchi_book_1d,gogolin_1dbook}.
This means that one has to keep track of all contributions to recover
the SU(2) invariance and, vice versa, neglecting irrelevant operators
{[}as we did to obtain the action in Eq.(\ref{S_total}){]} might
result in apparent inconsistencies in the formalism. In our case,
this problem has no consequences for our purposes because the operators
$O_{\pm}$ renormalize the couplings of the marginal contributions,
which we in any case we have neglected in relation to the relevant
contribution $\sim:\mathbf{N}_{1}\cdot\mathbf{N}_{2}:$. Therefore,
we will not consider these operators.

On the other hand, the first line in Eq. (\ref{eq:opeOKOK}) is physically
more interesting, as the operators $O_{3}$ and $O_{2c}$ were already
present in the action (\ref{eq:Sfixpoint}) corresponding to the Hubbard
model. If we insert (\ref{eq:opeOKOK}) into (\ref{eq:Zperturb}),
change variables as $\hat{r}=r-r^{\prime}$ and $R=\frac{r+r^{\prime}}{2}$
and integrate over $\hat{r}$ imposing a cutoff of order $a$, we
obtain an expression that renormalizes the first order contribution.
We identify the effective coupling for operators $O_{2c}$ and $O_{3}$
as: 
\begin{align}
\hat{G}_{2c}^{0} & =G_{2c}^{0}+\frac{3}{8}G_{K}^{2}
\end{align}
and the same for $\hat{G}_{3}^{0}$. Therefore, we have shown that
the interchain Kondo coupling generates an effective Hubbard repulsion
$U_{eff}=\frac{3a}{8}\frac{(J_{K}m_{2})^{2}}{\pi v_{F}}$ in the conduction
chain. The equation above can be physically interpreted as umklapp
processes (generated by integrating out fast spin fluctuations in
the Heisenberg chain at second order in the interchain Kondo coupling)
which mimic the effect of an interaction in the conduction band.

To determine the actual dependence of the charge gap $\Delta_{c}$
with respect to the parameters of the model, we need to derive the
RG flow equations. This is achieved following similar step as in the
previous paragraphs. The main idea is that the theory defined with
a microscopic cutoff $a$ should remain invariant under a scaling
transformation $a\rightarrow a\left(1+dl\right)$, where $dl$ is
a dimensionless infinitesimal. Therefore, the couplings $G_{\alpha}\left(a\right)$
in Eq. (\ref{eq:Zperturb}) must be changed in such a way that they
preserve the generating functional, i.e., $Z\left[a\right]=Z\left[a\left(1+dl\right)\right]$.
The method is standard and we refer the reader to Ref. \cite{cardy_book_renormalization}
for details. The renormalization group flow equations can be written
in terms of the coefficients $C_{\alpha\beta}^{\gamma}$ as 
\begin{align}
\frac{dG_{\gamma}}{dl} & =\left(2-\Delta_{\gamma}\right)G_{\gamma}-\pi\sum_{\alpha\beta}C_{\alpha\beta}^{\gamma}G_{\alpha}G_{\beta},\label{eq:generalRGeq}
\end{align}
where the coefficients $C_{KK}^{2c}=C_{KK}^{3}=-\frac{3}{4\pi}$ are
extracted from Eq. (\ref{eq:opeOKOK}). The remaining coefficients
are obtained by the OPEs between the corresponding operators. Following
the lines of Ref. \cite{delft_bosonization} we obtain straightforwardly:
\begin{align}
C_{2c\;3}^{3}=-\frac{1}{2\pi},\qquad & C_{3\;3}^{2c}=-\frac{1}{\pi}\nonumber \\
C_{2c\; K}^{K}=-\frac{1}{8\pi},\qquad & C_{3\; K}^{K}=-\frac{1}{2\pi}\label{eq:coefRG}
\end{align}
Inserting these values in Eq. (\ref{eq:generalRGeq}) we obtain Eqs.
(\ref{eq:RGG2c})-(\ref{eq:RGGK}) in the main text.

\end{document}